\documentclass[floatfix,preprint,aip,jap]{revtex4-2} 
\usepackage{lipsum}
\usepackage{graphicx}
\usepackage{amsmath}
\usepackage{amssymb}
\usepackage{epsfig}
\usepackage{subeqnarray}
\usepackage{color}
\usepackage{bbold}
\usepackage{dsfont}
\usepackage{tikz}
\usepackage{cancel}
\usetikzlibrary{positioning,calc}
\usepackage[version=3]{mhchem}
\usepackage{color}
\usepackage[normalem]{ulem}

\newcommand{\mt}[1]{\mathrm{#1}}

\newcommand{\fig}[1]{Figure~\ref{#1}}

\newcommand{\ie}{\textit{i.e.},~}

\newcommand{\eg}{\textit{e.g.,~}}

\newcommand{\mum}{~\mt{\mu m}}

\newcommand{\DGOS}{$\mt{D^2GOS}$~}
\graphicspath{{./Figures/}}

\begin{document}

\title{Detection of High Energy Ionizing Radiation using Deeply Depleted Graphene-Oxide-Semiconductor Junctions}
\author{Isaac Ruiz}
\author{Gyorgy Vizkelethy}
\author{Anthony E. McDonald}
\affiliation{Sandia National Laboratories, Albuquerque, United States of America}
\author{Stephen W. Howell}
\affiliation{Failure $\&$ Material Analysis Branch, Naval Surface Warfare Center Crane Division, Crane, United States of America}
\author{Paul M. Thelen}
\author{Michael D. Goldflam}
\affiliation{Sandia National Laboratories, Albuquerque, United States of America}
\author{Thomas E. Beechem}
\email{tbeechem@purdue.edu}
\affiliation{School of Mechanical Engineering and Birck Nanotechnology Center, Purdue University, West Lafayette, 47907, IN, USA}
\date{\today}

\begin{abstract}
Graphene’s linear band structure and two-dimensional density of states provide an implicit advantage for sensing charge.  Here, these advantages are leveraged in a deeply depleted graphene-oxide-semiconductor (\DGOS) junction detector architecture to sense carriers created by ionizing radiation. Specifically, the room temperature response of a silicon-based \DGOS junction is analyzed during irradiation with 20 MeV $\mt{Si^{4+}}$ ions. Detection was demonstrated for doses ranging from 12-1200 ions with device functionality maintained with no substantive degradation. To understand the device response, \DGOS pixels were characterized post-irradiation via a combination of electrical characterization, Raman spectroscopy, and photocurrent mapping. This combined characterization methodology underscores the lack of discernible damage caused by irradiation to the graphene while highlighting the nature of interactions between the incident ions and the silicon absorber. 
\end{abstract}
\maketitle

\section{Introduction}
Compared to traditional ionization chambers and scintillation detectors, solid-state radiation detectors have several advantages due to their smaller size, lower power consumption, higher atomic density, and higher ionization energy conversion efficiency.\cite{milbrath_2008,lutz_1999,darambara_2002} Furthermore, with access to materials with a wide range of physical properties, solid state detectors can be tailored through selection of material atomic number, bandgap, and density, making them suitable for a range of applications.\cite{owens_2004} However, existing implementations of these detectors suffer from several disadvantages. Specifically, there remains a dearth of single-crystal high-Z materials suitable for the task that can be synthesized cost-effectively at scale.  Meanwhile, more common low-Z materials are predisposed to ion-induced degradation and oftentimes necessitate cooling (\eg narrow bandgap semiconductors).\cite{goiffon_2012,lee_2015b,delsordo_2009,llacer_1972} Regardless of absorber choice, external signal amplification is typically required for sensing low ion doses. These deficiencies are implicit to the materials and architectures traditionally employed in radiation detection. Overcoming these limitations, therefore, demands alternative materials and architectures. 

The deeply depleted graphene-oxide semiconductor (\DGOS) detector provides such an alternative architecture. The \DGOS relies on graphene to capacitively sense charge that collects within a potential well formed at the interface between a dielectric and deeply depleted semiconductor. It therefore removes the need to create and extract charge from a single medium giving flexibility in the choice of semiconductor.\cite{foxe_2009,childres_2010,foxe_2012a,patil_2011,koybasi_2013,cazalas_2016}  Previous reports have shown the architecture to be an extremely sensitive photodetector.  For example, \DGOS architectures have shown signal-to-noise ratios (SNR) $>100$ and optical responsivities more than 25,000 A/W in sensing electromagnetic radiation from the ultraviolet (UV) to near infrared (NIR).\cite{lee_2003,howell_2017,ruiz_2019} As any charge collecting within the depletion well is mirrored in graphene regardless of its origins, the architecture should be effective for radiation sensing as well.

The sensitivity of the \DGOS stems from graphene’s characteristics as a near-idyllic charge sensing layer.  Its single-atom thickness and the low-Z of carbon makes it nearly transparent to incident radiation.\cite{alexandrou_2016} Simultaneously, its linear band structure and two-dimensional density of states lead to a low intrinsic carrier concentration. This low intrinsic carrier concentration means that a single charge added to the population yields a large relative change in carrier density, which can be monitored through measurements of current flowing through graphene at a constant voltage.  Lastly, graphene’s high mobility acts as a gain mechanism for this current and thus a built-in amplifier.\cite{blake_2007,adam_2007,banszerus_2015}  

Despite the benefits inherent to graphene, graphene-based devices can be irreversibly altered upon exposure to radiation calling into question their utility as a radiation sensor.\cite{alexandrou_2016,cazalas_2019}  For example, gamma exposure can reduce graphene-based transistor performance as can X-ray, or ion irradiation.\cite{cazalas_2019,lee_2015b,francis_2014}  However, it remains unclear, whether these changes occur owing to damage within the graphene or the surrounding device structure (\ie substrate, dielectric, passivation layers).  The former suggests a fundamental constraint for the use of graphene as a radiation sensor whereas the latter can be overcome with the traditional means of radiation hardening developed for microelectronics in general.

With this dual motivation, we examine here the utility of a \DGOS operating in deep depletion as an avenue to particle radiation detection and as a tool to assess where radiation damage is most likely to occur in graphene devices. Practically, we employ a Si-based \DGOS junction to demonstrate room temperature sensing of 20 MeV $\mt{Si^{4+}}$ ion pulses ranging from 10 ms to 1 s in duration with a total number of ions varying from 12 to 1200 ions/pulse.  Post-irradiation, the device was characterized via a combination of Raman spectroscopic and photocurrent imaging with results interpreted using context provided by Monte Carlo modeling of the irradiation.  While irradiation does induce irreversible changes in the device on the whole, the graphene itself is not substantively altered. The graphene device is changed by irradiation but not the graphene itself. Unlike typical solid state detectors, the \DGOS junction allows for: (1) the accumulation of radiation induced charge near the graphene, (2) direct readout of the induced charge, and (3) the rapid dissipation of accumulated charge, by operating in and out of depletion.  

\section{Materials and Methods}
\begin{figure}[htbp]
\centering
\includegraphics[width=85 mm]{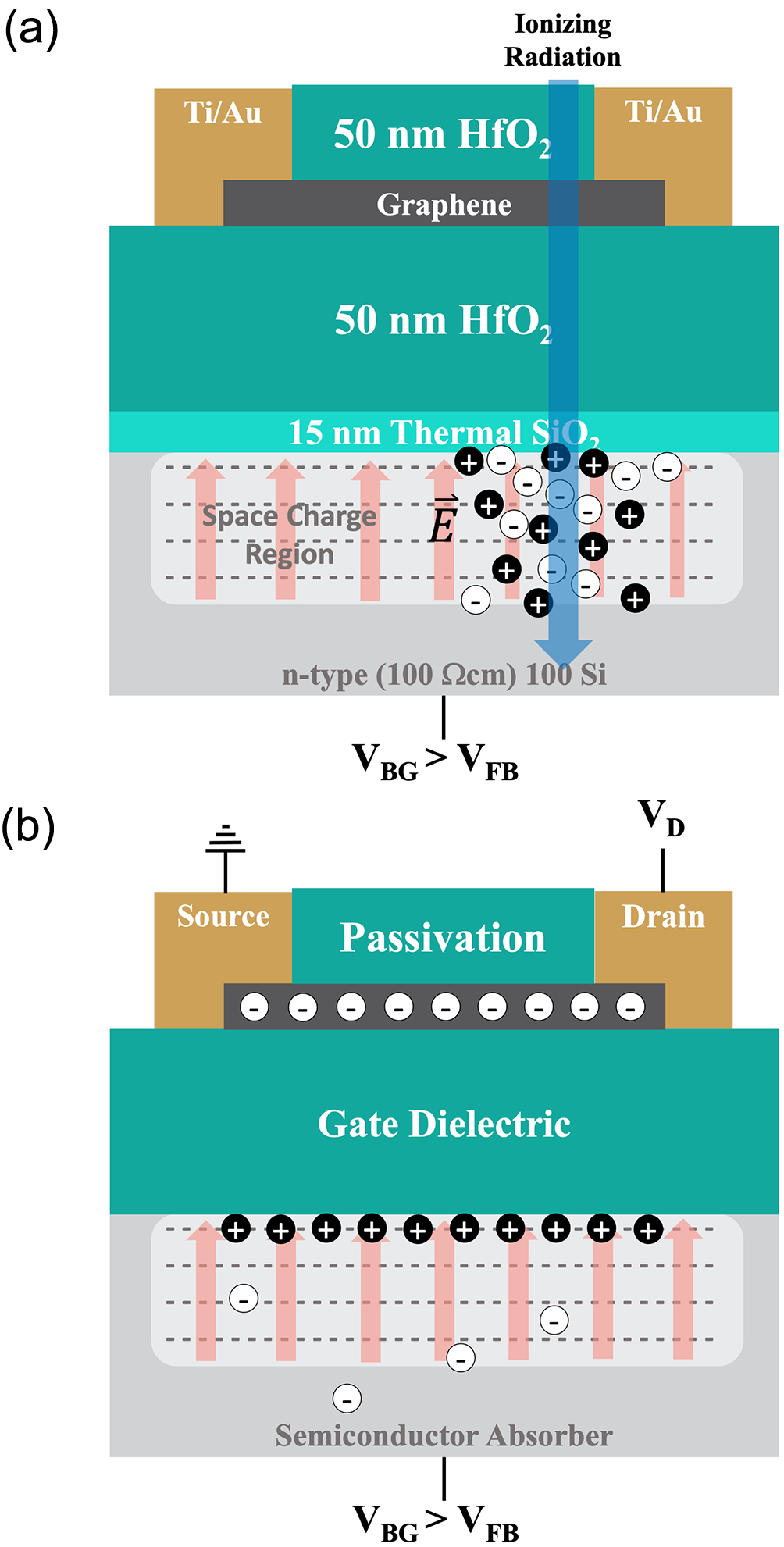}
\caption{Cross sectional schematic of a \DGOS junction with an n-type substrate as it is biased into deep depletion ($V_{BG}$ $>$ $V_{FB}$) and (a) electron-hole pairs are created by radiation within the space charge region. The generated electron-hole pairs are then (b) separated by the built-in electric-field causing holes to accumulate at the semiconductor-oxide interface thereby inducing the opposite charge in the graphene channel.}
\label{Fig_1}
\end{figure}

The \DGOS junction was fabricated on a low-doped (100 $\mt{\Omega cm}$) n-type (100) Si substrate, with a 15 nm thermal oxide passivation layer grown on the surface to assure a low interface defect density. Shown schematically in \fig{Fig_1}, devices consisted of a graphene channel 200 x 250 $\mum$ in size fabricated via processes described in detail previously.\cite{ruiz_2019} Characterization of the \DGOS junction’s response to $\mt{Si^{4+}}$ ions was performed in-situ at Sandia National Laboratories’ Ion Beam Laboratory using the microbeam of the HVE 6 MV Tandem accelerator. The in-situ nature of the experiment made it impossible to prevent all extraneous light from reaching the device.  This caused a small, but non-negligible, photocurrent impacting the ion sensing measurements, as is discussed later.

The devices were irradiated with 20 MeV $\mt{Si^{4+}}$ ions at a rate of 1200 ions/s for durations of 10 ms to 1 s.  Irradiation occurred in two 50×50 $\mum$ locations both within, and just outside, the graphene-covered area of the device, as delineated by pink boxes in the inset of Figure 2b. Data presented in this manuscript describes effects stemming from irradiation within the active area.

Ion sensing was accomplished by first biasing the junction into accumulation ($V_{BG}$ $<$ $V_{FB}$) for 2 s followed by a rapid biasing into deep depletion ($V_{BG}$ = 4.0 V) for 20 s. The ion pulse was triggered during this process resulting in a stream of $\mt{Si^{4+}}$ ions impacting the device about 1 s after deep depletion occurred. Throughout these steps, the source-drain current through graphene was monitored at a constant $V_{D}$=0.1 V drain voltage to assess the device’s real-time response to the irradiation. 

Post-irradiation Raman and photocurrent imaging were implemented using a Witec alpha300R system equipped with 532 and 785 nm lasers each focused to a diffraction limited spot with a 50X/0.55 NA objective.  For Raman imaging, a 300 x 300 $\mum$ region across the device was interrogated with spectral acquisitions separated by 2 µm.  Two separate data sets were collected to investigate the graphene and silicon separately.  For the graphene, the 532 nm laser was employed for higher surface sensitivity with collected light dispersed onto a 600 l/mm grating providing spectral accuracy to within 1 $\mt{cm^{-1}}$.  Silicon was probed using the 785 nm laser.  At this wavelength, the skin depth of silicon is $\sim 9 \mum$ thereby increasing light collection from the region most affected by the ions.\cite{palik_1998}  Raman scattering at these longer wavelengths was analyzed with a separate spectrometer optimized for NIR operation and dispersed with a 1200 l/mm grating allowing for specification of the relative silicon peak position to less than 0.01 $\mt{cm^{-1}}$.  

\section{Results and Discussion}
\fig{Fig_1} schematically describes the basic operational principles of the device.  In short, a graphene-oxide-semiconductor (GOS) junction is biased into deep depletion through the sudden application of a back-gate bias, $V_{BG}$.  A space-charge region is created near the semiconductor-oxide interface (see \fig{Fig_1}(a) for an n-type semiconductor) when $V_{BG}$ is greater than the flat band voltage, $V_{FB}$. With interaction of the radiation with the semiconductor, electron-hole pairs created within the space-charge region are separated by the built-in electric field with minimal recombination of the minority carriers. For an n-type absorber, holes accumulate at the semiconductor interface inducing opposite charges (electrons) in the graphene channel.  The change in carrier concentration within the graphene channel alters its resistivity. These changes to graphene are read-out by monitoring the current between the source and drain contacts ($I_D$) at a constant bias ($V_D$) as shown in \fig{Fig_1}(b).  Current will continually be modified until charge fills the potential well causing saturation.

\begin{figure}[htbp]
\centering
\includegraphics[width=150 mm]{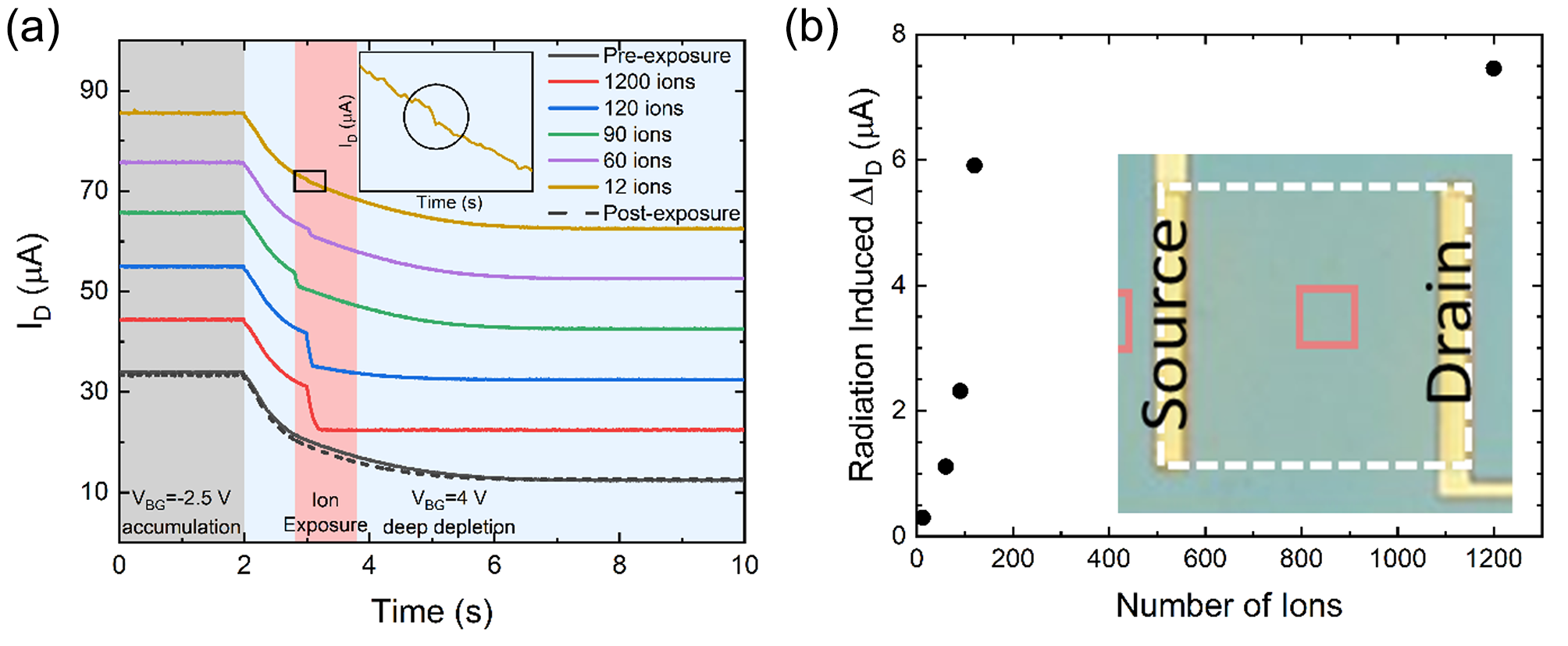}
\caption{$I_D$(time) response of the same \DGOS junction pixel when repeatedly exposed directly to 20 MeV $\mt{Si^{4+}}$ ions. Curves obtained during irradiation in (a) are offset by 10 $\mu A$ for clarity. The inset presents an enlarged view of the data contained within the small box that overlays the time duration of the 12 ion pulse. (b) Magnitude of ion-generated (\ie radiation induced) current, $\Delta I_D$. }
\label{Fig_2}
\end{figure}

The \DGOS junction is capable of sensing the 20 MeV $\mt{Si^{4+}}$ ions, as is evident upon examination of \fig{Fig_2}.  This is seen by the sudden jumps in current through the graphene observed when ions impinge upon the device (pink shaded region in \fig{Fig_2}(a)).  Qualitatively, the magnitude of the jumps scale with the ion dose. For a more quantitative examination of ion sensitivity, we extract the radiation induced change in current ($\Delta I_D$) obtained by taking the difference in current immediately before and after the ion pulse. The radiation induced $\Delta I_D$ is shown in \fig{Fig_2}(b) with detection demonstrated for irradiation of 12-1200 ions.

Next, we more closely examine the temporal evolution of current in the \DGOS as shown in \fig{Fig_2}.  Upon biasing into deep depletion (t = 2 s in \fig{Fig_2}(b)), the current through the graphene changes accordingly. Apart from the presence of ions, the current would ideally remain constant after this initial change as the gate voltage is held constant. In reality, a continual reduction is observed with eventual current saturation reached 7 s after depletion biasing, as shown by the pre-exposure (bottom/black curve of \fig{Fig_2}(a)).  This indicates that charge is collecting at the silicon/oxide interface independent of the ions and ultimately filling the depletion well.  Since similar devices measured in a separate fully dark enclosure required 60 s to saturate, we deduce that stray light is the primary cause for the time dependent decrease in current for the unirradiated response.  Consequently, ion-generated charge will compete with the charge generated by stray light in filling the potential well.   While this will reduce the present device’s radiation sensitivity, the problem is easily circumvented in practice via suitable optical filters placed in front of, or directly integrated atop of, the radiation detector. 

Detection of 12 ions is not a fundamental limit to the sensitivity of these devices. Rather, it is simply a limit of the experimental configuration in which stray light and experimental noise compete with the ion induce charge generation. Fundamentally, the limit is dictated by the fluctuation of current within the graphene channel in comparison to the current induced by capacitively coupled charge created by the ion.  Based on the variation in current in accumulation at constant applied bias (t$<$2 s), conditions under which we would expect to measure a constant current, the noise level of the device/characterization system is estimate to be 0.3$\%$ as calculated from the ratio of the standard deviation of the current to its average level. Based on this noise level, in combination with a linear extrapolation of the radiation induced $I_D$ as function of number of ions, we estimate that the noise floor of our device/characterization setup is approximately 3 ions. Again, this does not represent a fundamental limit of this device architecture but is rather a limit of the present device and characterization setup. 

The maximum number of ions that can be sensed, meanwhile, is limited by the amount of charge required to fill the potential well.  This is seen directly when examining the trace captured during dosing of 1200 ions where saturation is apparent even before the ion dose ends.  Thus, not all the ion generated charge is transduced by the graphene.  This is apparent in \fig{Fig_2}(b) where the radiation induced current clearly diverges from its initial linear trend at this high dosing. Regardless, the data show the capability of the \DGOS architecture to sense radiation via transduction of charge created in an adjacent deeply-depleted semiconductor.

It is therefore necessary to assess whether the ions being sensed appreciably harm the device in a manner that would preclude any practical utilization to this end.  For this reason, we first qualitatively evaluate cumulative damage from 20 MeV ion bombardment via measurement of the post-exposure current response. Only slight differences between pre- and post-ion current are observed (see \fig{Fig_2}(a) ). The consistent slopes of the pre- and post-exposure curves indicate that the mobility of the graphene has not appreciably changed.  Post-irradiation, the device is therefore still functional, and any damage has only a negligible effect on the basic performance of the device. Use of the atomically thin graphene does not necessarily preclude operation in this modality even though graphene devices are known to be irreversibly altered by radiation.\cite{alexandrou_2016}  

This conclusion is supported by examining the effect of ions on the device structure via Raman spectroscopic imaging.   Raman images indicate that silicon is damaged, but the graphene is\textemdash to within the limits of our measurement system\textemdash unaffected by the exposure.  The false color image of Figure 3(a), for example, highlights softening of the silicon in regions exposed to ion irradiation as indicated by the reduction in peak position.  Similar contrast is not observed, however, in the defect-monitoring I(D)/I(G) ratio of the graphene (Figure 3(b)) within the irradiated zone. Taken together with lack of change in the graphene mobility, this indicates that silicon is affected more by the ions than the graphene.  The atomically thin material is not the weakest link.    

\begin{figure}[htbp]
\centering
\includegraphics[width=150 mm]{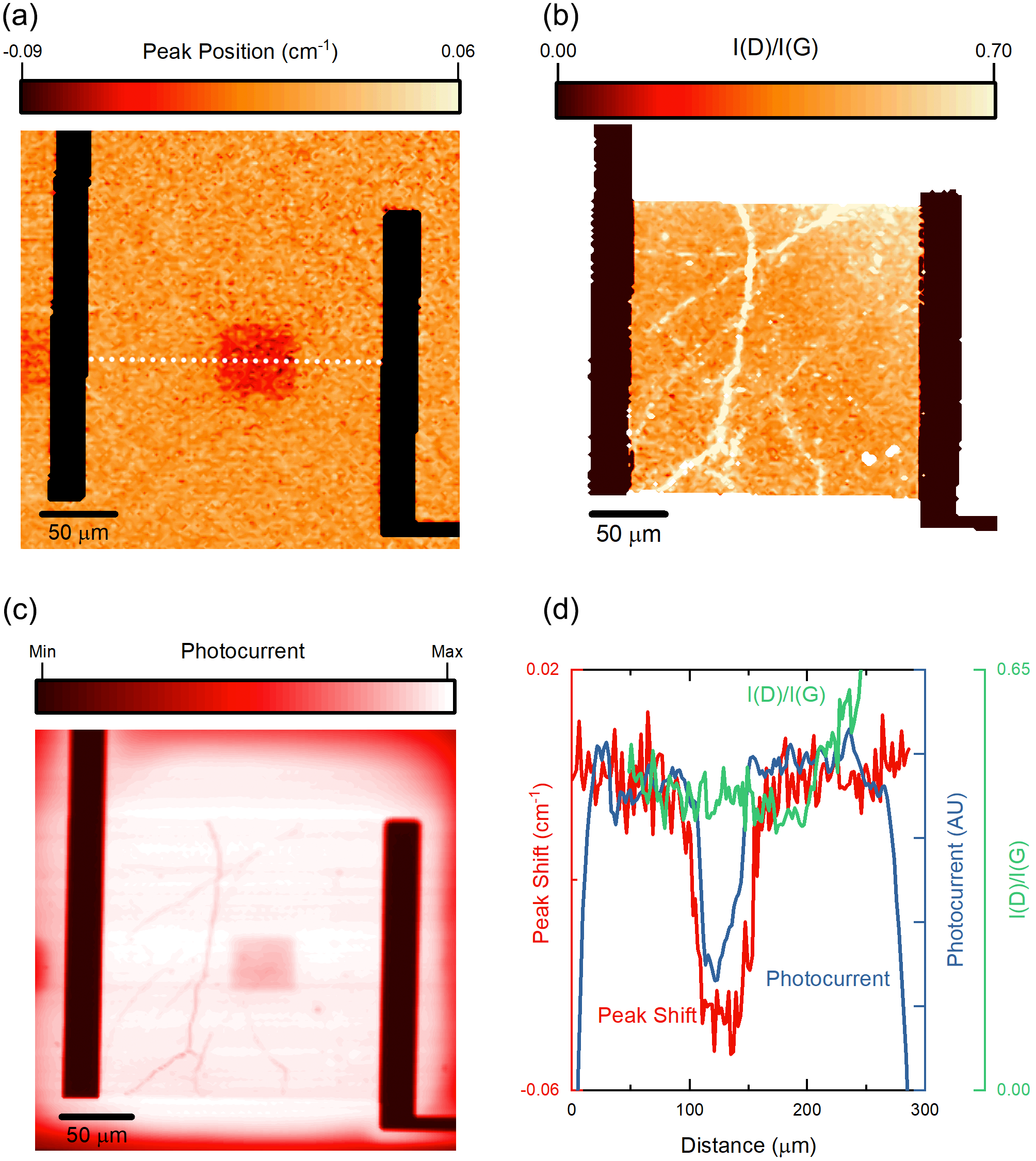}
\caption{Raman images highlighting softening of the (a) silicon within region of irradiation and the negligible change that occurs in the (b) graphene as evidenced by its unaffected I(D)/I(G) ratio.  (c) Photocurrent reduces within the irradiated region.  This could be caused by increased recombination due to damage in either the silicon itself or at the interface between the silicon and \ce{HfO_2}. (d) Cross sectional profiles along the dotted line of (a) corresponding to the Si peak-position, I(D)/I(G) ratio, and photocurrent.}
\label{Fig_3}
\end{figure}

\begin{figure}[htbp]
\centering
\includegraphics[width=140 mm]{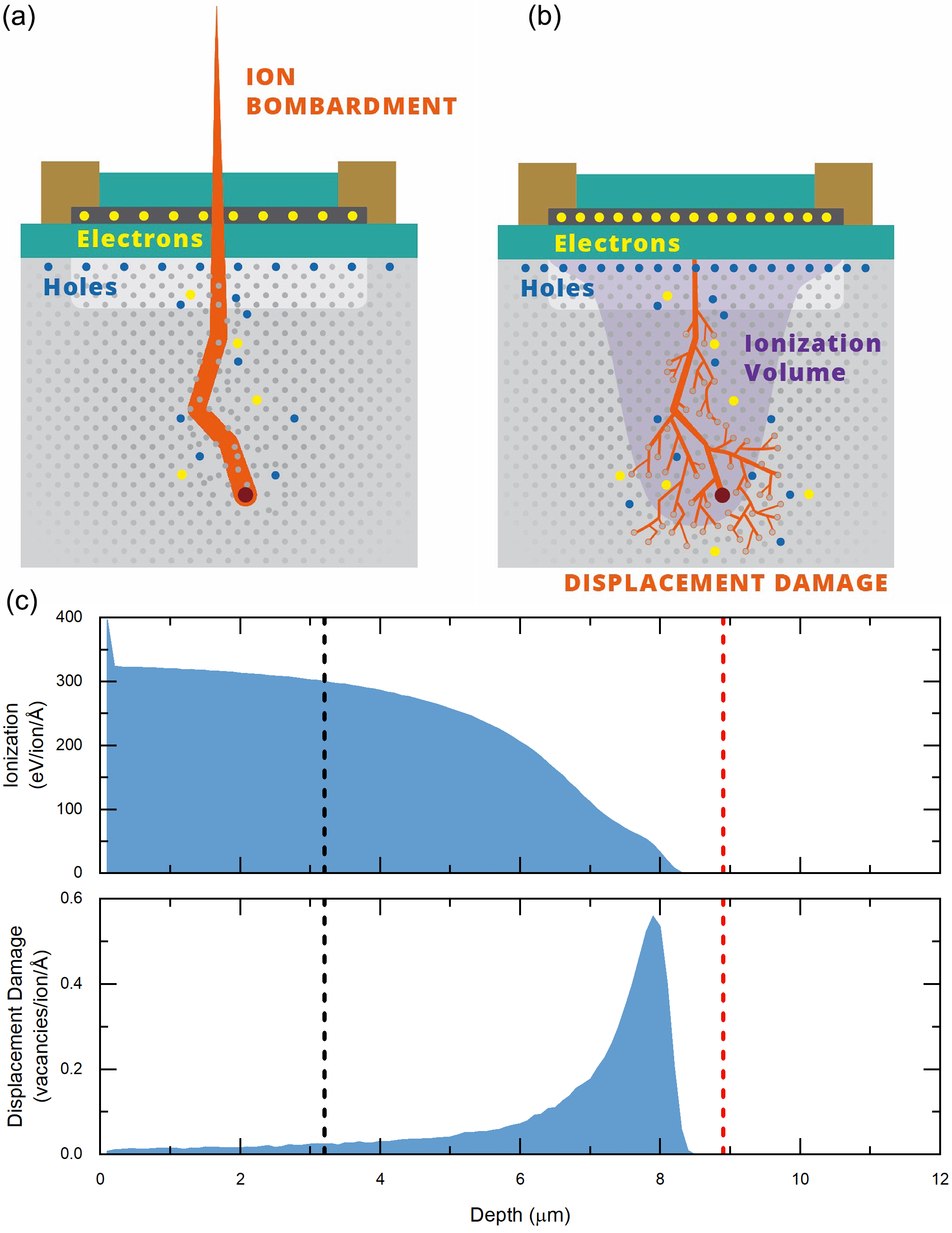}
\caption{Schematic representation of where ion induced (a) elastic (nuclear stopping) and (b) inelastic (electronic stopping) interactions preferentially occur within a \DGOS device. (c) Ionization and displacement damage profiles calculated via SRIM. Vertical red and black lines correspond to the skin depth of 785 nm light\cite{palik_1998} and the nominal depletion width of the device, respectively.}
\label{Fig_4}
\end{figure}

This result can be understood when considering how ions interact with the device structure using the schematic illustrations of \fig{Fig_4}(a,b) as a guide.  An incoming ion penetrates through the outer device layers (thin dielectrics and graphene) and the Si, losing energy through elastic (nuclear stopping) and inelastic (electronic stopping) interactions before coming to rest within the crystal.  These interactions create an ionization volume within which electron-hole pairs are created. In addition, the impinging ions create displacement damage localized within the bulk silicon crystal at a depth dependent upon the energy of the ion.  Lower energy ions would likely have a stronger influence on device performance than that of 20 MeV ions. Specifically, these ions have a higher likelihood of inducing damage within the device space charge region, or directly within the graphene. Future use of such devices for radiation detection would require examination of response with reduced ion energy.  Ion energy dependence is beyond the scope of this study, however, which is instead centered on demonstrating proof of principle.  

To be more quantitative, ionization and displacement damage were quantified via Stopping Range of Ions in Matter (SRIM) simulations as shown in \fig{Fig_4}(c). These simulations indicate that ionization is largest near the Si surface but persists to nearly 8 $\mum$ below the boundary.  Displacement damage is localized at this 8 $\mum$ depth as well, which is more than double the thickness of the space charge region created in these devices ($\sim$3.2 $\mum$, black dashed line in \fig{Fig_4}(c)). Charge is primarily collected within the space charge region, which is encompasses much of the ionization volume. The overlap of the space charge region and the ionization volume underpins the ion sensitivity of the \DGOS.  Displacement damage, meanwhile, is comparatively inconsequential to performance.  This is because displacement damage occurs beyond the depletion depth of the device.  Thus, even though displacement damage increases e-h recombination, the \DGOS device does not ``feel" it because it is not collecting charge from the displacement affected volume. 

This does not suggest that the displacement damage has no effect.  Photocurrent is markedly reduced in the regions of ion exposure as shown in \fig{Fig_3}(c).  A reduction of photocurrent indicates a lower photogenerated carrier density at the Si-\ce{SiO_2} interface and, therefore, a smaller shift in the current through graphene. As displacement damage in the Si substrate is linked to an increase in carrier trapping and recombination rates.\cite{xapsos_1987} , we postulate that the increased defect density within the depletion region induced by 20 MeV ions, as demonstrated by the softening of Si from our Raman measurements (\fig{Fig_2}(a) ), results in a decrease in charge collection efficiency. 

Alternative mechanisms for the photocurrent reduction could also be at play, however.  For example, damage alters the optical properties of the stack and thus light absorption could be reduced as evidenced by the nominal change in reflectance observed via monitoring of the Raman signal’s baseline (not shown).  Additionally, radiation induced traps at the oxide/silicon interface may also reduce the photocurrent.  

Regardless of causation, the reduction in photocurrent within the irradiated region accompanied by the softening of silicon point to the substrate being most acutely impacted by the ions.  This is further supported by the constancy in the I(D)/I(G) (\ie defect) ratio of the graphene and the relatively constant mobility even after irradiation.  We therefore conclude that not only does the \DGOS present a viable radiation detector but that the graphene is comparatively insensitive to the irradiation relative to the most common of semiconductors. This supposition should next be further analyzed with higher-Z absorbers that are implicitly more resilient to radiation damage. 


\section{Conclusions}
Here, we have demonstrated that the \DGOS junction is sensitive to Si ion irradiation with the ability to detect as few as twelve 20 MeV ions. While the silicon absorber examined here is slightly damaged by irradiation, damage to the graphene is minimal and below sensitivity level that can be detected with our Raman system.  As the detector architecture itself is absorber agnostic, higher-Z semiconductors provide a path forward leveraging the advantages of the \DGOS for sensitive and robust radiation detection that can be tailored for specific scenarios. 


\section{Acknowledgements}
The authors thank the Center for Integrated Nanotechnology (CINT) for access to their atomic layer (ALD) deposition system and their electron beam evaporator, as well as Ava Howell for her help with the well-filling measurements. Sandia National Laboratories is a multimission laboratory managed and operated by National Technology and Engineering Solutions of Sandia, LLC, a wholly owned subsidiary of Honeywell International Inc., for the U.S. Department of Energy’s National Nuclear Security Administration under contract DE-NA0003525. This work was supported by the Laboratory Directed Research and Development (LDRD) program at Sandia National Laboratories, and was performed, in part, at the Center for Integrated Nanotechnologies, a U.S. DOE, Office of Basic Energy Sciences user facility. This paper describes objective technical results and analysis. Any subjective views or opinions that might be expressed in the paper do not necessarily represent the views of the U.S. Department of Energy or the United States Government.

\section{Data Availability}
The data that support the findings of this study are available from the corresponding author upon reasonable request.

\bibliography{Library_Updated}

\end{document}